# A Use Case-Engineering Resources Taxonomy for Analytical Spreadsheet Models

Thomas A. Grossman and Vijay Mehrotra

University of San Francisco, 2130 Fulton St., San Francisco CA 94117 USA

tagrossman@usfca.edu, vmehrotra@usfca.edu

**ABSTRACT**

This paper presents a taxonomy for analytical spreadsheet models. It considers both the use case that a spreadsheet is meant to serve, and the engineering resources devoted to its development. We extend a previous three-type taxonomy, to identify nine types of spreadsheet models, that encompass the many analytical spreadsheet models seen in the literature. We connect disparate research literature to distinguish between an "analytical solution" and an "industrial-quality analytical spreadsheet model". We explore the nature of each of the nine types, propose definitions for some, relate them to the literature, and hypothesize on how they might arise. The taxonomy aids in identifying where various spreadsheet development guidelines are most useful, provides a lens for viewing spreadsheet errors and risk, and offers a structure for understanding how spreadsheets change over time. This taxonomy opens the door to many interesting research questions, including refinements to itself.

## 1. INTRODUCTION

It is well-understood that spreadsheets are indispensable to business (e.g., Croll 2007, Grossman, Mehrotra, and Özlük 2007), but can also be a source of risk and costly errors (e.g., EuSpRIG 2023, Panko 1998, but also Powell, Baker, and Lawson 2008a, 2008b, 2009a, 2009b). Researchers and practitioners have long been working on articulating standards and practices to reduce risk and error, and also to increase the productivity of spreadsheet programmers, and the effectiveness of spreadsheet users in organizations.

Because of the scope and complexity of the spreadsheet space, any professional practice recommendations must be limited to a well-defined "domain" or class of spreadsheets (Grossman, Mehrotra, and Sander 2011, henceforth referred to as "GMS"). This leads naturally into the need for a taxonomy (set of named classifications) for spreadsheets. GMS consider the domain of analytical spreadsheet models, and provide a taxonomy based on how the spreadsheet is used.

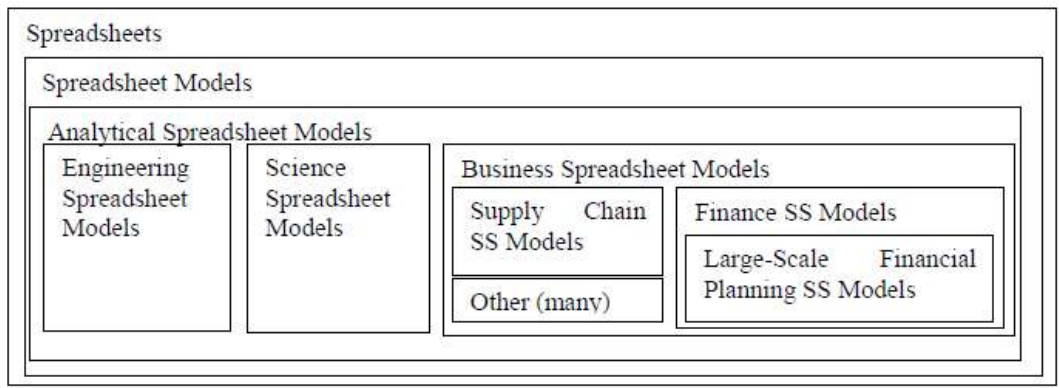

*Figure 1: The Analytical Spreadsheet Model Domain (Grossman, Mehrotra and Sander 2011)*

The domain of analytical spreadsheet models (see Figure 1) is distinct from spreadsheet models in general, and from spreadsheets in general. GMS provide a five-part definition of an *analytical spreadsheet model* as a spreadsheet computer program, that implements a mathematical model, for purposes of analysis, that serves as an organizational asset, employed in an important context.

GMS indicate that analytical spreadsheet models are primarily about the business logic embedded in the model, and that models that are primarily data-driven with a large dataset and relatively few cell formulas do not fall in this domain.

GMS differentiate among three types of analytical spreadsheet model, namely "personal productivity tool" (something "quick and dirty" and relatively unimportant); "analytical spreadsheet model" (which is used regularly by someone other than the developer), and "spreadsheet application" (that is deployed to multiple less-sophisticated users). They briefly suggest that the recommended amount of spreadsheet engineering effort is higher for each type of respective spreadsheet. GMS then move on and focus on the characteristics and guidelines for "high quality" analytical spreadsheet models, and do not develop the taxonomy.

This paper extends, refines, and formalizes the three-element taxonomy of Grossman, Mehrotra, and Sander 2011 for analytical spreadsheet models into a richer nine-element model that provides significant insight into how we can think about analytical spreadsheet models, how they are created and should be created, and how they are changed over time.

**1.1 Contribution**

This paper presents a new framework, or "taxonomy" for understanding analytical spreadsheet models. It examines the relationship between the *use case* that a spreadsheet is meant to serve, and the *engineering resources* devoted to the spreadsheet's development. The taxonomy identifies and proposes a name for nine types of analytical spreadsheet model, extending and refining a previous three-type taxonomy. We identify a new distinction between two previously-merged types of spreadsheets, which are the "analytical solution" and the "industrial-quality analytical spreadsheet model".

We explore the nature and knowledge of each spreadsheet type, including criteria for a spreadsheet model being considered that type, the appearance of that type in the literature, and how that type of spreadsheets might arise in practice. This taxonomy allows us to examine with greater insight (and even with sympathy) the phenomena of both under- and over-engineered spreadsheets and the costs and risks associated with them.

We believe that spreadsheet development guidelines are not "one size fits all" but should instead vary depending on the business context. This taxonomy provides a framework to understand existing development recommendations, and provides a map to spreadsheet types where development recommendations or risk management measures might usefully be articulated.

We anticipate that this taxonomy can help us understand the evolution and risk of an important analytical spreadsheet model, by providing a structure to explain how a spreadsheet's use changes over time, with or without adequate planning and investment to support that change.

We emphasize that this research is preliminary. We anticipate that there will be refinements and alterations to various aspects of the taxonomy. Much work remains to be done to fully flesh out the details and criteria for each type of spreadsheet, and to test various insights and hypotheses that arise.

## 1.2 Overview

In Section 2 we critically examine and refine the GMS three-type taxonomy for analytical spreadsheet models, and clear up some confusing elements from the original. We refine and extend it to two dimensions, and place the original three types into the larger two-dimensional structure. The new taxonomy considers the intersection of three use cases, discussed in Section 3, and three engineering resource levels (Section 4). In Section 5, we combine the use cases and engineering resource levels to yield a new two-dimensional taxonomy featuring nine types of analytical spreadsheet model, which we name and explore in detail. We conclude in Section 6 with suggestions for further research.

## 2. REFINING AND EXTENDING THE GMS THREE-TYPE TAXONOMY

In this section we examine the GMS three-type taxonomy. We critique this taxonomy, clarify and refine it, and extend it from one dimension to two. This prepares us to develop the new use case-engineering resources taxonomy in Sections 3, 4, and 5.

The GMS taxonomy for analytical spreadsheet models, which admits three types of spreadsheet model, is formalized in Figure 2.

| Personal Productivity Tool (PPT) | Analytical Spreadsheet Model (ASM) | Spreadsheet Application (App) |
|---|---|---|

**Type of Analytical Spreadsheet Model**

*Figure 2: Three-Type Taxonomy (Adapted from Grossman, Mehrotra, and Sander 2011)*

The GMS taxonomy considers three types of spreadsheet, the Personal Productivity Tool, Analytical Spreadsheet Model, and Spreadsheet Application. We begin by clarifying the terminology.

### 2.1 Clarifying Terminology and Revisiting a Key Assumption

In the GMS paper, the term "analytical spreadsheet model" is used in three different ways: (A) the domain of spreadsheet model under consideration (see Figure 1); (B) as the classification of spreadsheet model (i.e., the horizontal axis in Figure 2); and (C) as the name of a cell in Figure 2 ("Analytical Spreadsheet Model (ASM)").

In this paper, we use "analytical spreadsheet model" only for (A), the *domain*. (The domain of analytical spreadsheet model is defined in Section 1.) This paper (like GMS) considers only spreadsheets that are analytical spreadsheet models. We replace the other uses of "Analytical Spreadsheet Model" with more appropriate terminology in Section 2.2.

Next, we relax a key assumption in GMS. GMS conflate the level of "spreadsheet engineering investment" and the Type of Analytical Spreadsheet Model. They briefly indicate that personal productivity tools should receive lesser spreadsheet engineering investment, analytical spreadsheet models more investment, and spreadsheet applications even more investment. That is, GMS assume alignment between type of model and the investment to create that model.

However, although real-world development projects (including building spreadsheets) sometimes do receive the "just right" level of investment, they can also be provided with insufficient or even excess investment. (cf. Southey 1837). Therefore, we extend in Figure 3 the taxonomy of Figure 2 to explicitly represent varying investment levels.

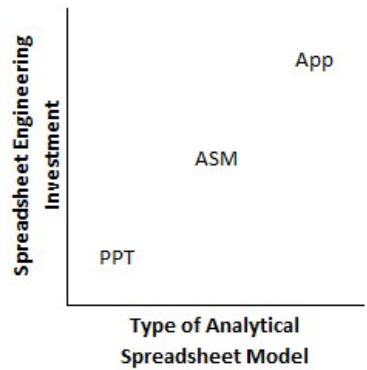

*Figure 3: GMS Taxonomy Extended to Display the Different Investment Levels, Indicating the Assumption of Alignment of Spreadsheet Engineering Investment and Model Type*

The three spreadsheet model types from Figure 2 appear in Figure 3, showing the GMS alignment between spreadsheet engineering investment and type of analytical spreadsheet model

**2.2 Refined Terminology: Use Case and Engineering Resource Level**

Next, we refine the axes of Figure 3. We rename the horizontal axis from "Type of Analytical Spreadsheet Model" to "Use Case" which is more descriptive. We use Google's Oxford Languages definition of *use case* to mean "a specific situation in which a product or service could potentially be used". We define three use cases, called "Unique Analysis", "Business Process-Embedded", and "Spreadsheet Application". (These three use cases are discussed in detail in Section 3.)

We rename the vertical axis from "Spreadsheet Engineering Investment" to "Engineering Resources" which is shorter and encompasses resources such as time and managerial attention in addition to expense. We indicate three levels of engineering resources, called "Low", "Medium", and "High". (These three levels of engineering resources are discussed in detail in Section 4.)

This yields the Use Case-Engineering Resources Taxonomy (Figure 4). This taxonomy allows us to consider the interaction between the engineering resources deployed to build an analytical spreadsheet model, and the situation in which that model will be used. We add gridlines to highlight that there are nine combinations of Engineering Resources and Use Case.

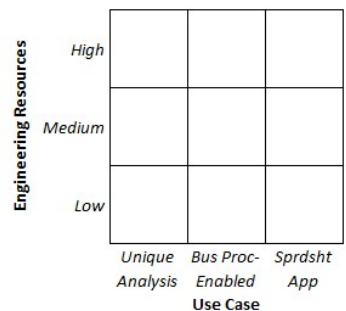

*Figure 4: The Use Case-Engineering Resources Taxonomy*

Let us now revisit and refine the three types of spreadsheet from GMS in Figures 2 and 3. We rename them, and place them into Figure 5 in boxes that indicate alignment between Engineering Resources and Use Case. We adjust the name of the bottom left box to "Personal Productivity Spreadsheet" (PPS) rather than "Personal Productivity Tool" because "tool" is too restrictive, and because "tool" can be assigned a different meaning (e.g. Olavson and Fry 2008). The center box is called "Industrial-Quality Analytical Spreadsheet Model" (IQASM) to emphasize that it has received a goodly level of investment. The top right box is called "Planned Application" to emphasize that an appropriate level of engineering resources are used in its development. These refinements are shown in Figure 5.

*Figure 5: The Use Case-Engineering Resources Taxonomy Showing the Three Types of Spreadsheet Models That Have Alignment Between Use Case and Engineering Resources*

### 2.3 Alignment, Waste, and Danger?

It is tempting (and implicit in GMS) to assume that Engineering Resources should be aligned with Use Case. If this is true, then our taxonomy boils down to Figure 6, where situations with Engineering Resources higher than necessary represent waste, and situations with Engineering Resources lower than necessary represent danger.

*Figure 6: Simplistic Use of the Taxonomy that Assumes Alignment Between Engineering Resources and Use Case is Always Appropriate*

However, upon further examination in Section 5, we shall see that this is not the case, and that it can be appropriate for a spreadsheet to reside in a misaligned cell (off the up-right diagonal). But first, let us further develop the taxonomy.

### 3. THE THREE USE CASES

We now provide more detail on the use-case dimension of the taxonomy in Figure 5. There are three use cases: Unique Analysis, Business-Process Enabled, and Spreadsheet Application. (Disclaimer: It is our hope that these three use cases and aspects encapsulate the bulk of important real-world spreadsheet models – and therefore are useful – but we are not quite prepared to claim that they capture every possible circumstance.) Examples of spreadsheets from each use case can be found in Grossman, Mehrotra, and Özlük (2007) and elsewhere in the literature.

The <u>Unique Analysis</u> use case is where a spreadsheet model is built to undertake a single calculation or to provide a model used for analysis in a situation that is limited in scope or time. The spreadsheet is not shared (expect perhaps within a "tight team" that collaborates well and respects spreadsheet integrity). For example, a financial model for a particular project, or the author's mortgage interest rate spreadsheet.

The <u>Business-Process Enabled</u> use case is where a spreadsheet model is used on a regular or routine basis as part of an established business process. Typically, the model is transferred to and used by the business process owner (and is unlikely to be relevant to anybody else). The nature of these spreadsheets is that they are necessarily further transferred over time when the user changes jobs. For example, a spreadsheet to determine weekly orders from multiple vendors; a monthly budget spreadsheet; a spreadsheet that summarizes cumulative capital investment. This use case seems to be the source of the "Accidental Legacy Systems" observed by Grossman, Mehrotra, and Özlük (2007).

The <u>Spreadsheet Application</u> use case is where a spreadsheet model is written by a developer and distributed to other people (with an unspecified level of intentionality and planning), resulting in heavy usage of the spreadsheet by multiple users.

In Table 1, we describe four aspects of each use case: intended frequency of use; who can access the spreadsheet; number of users; and reliability of user behavior.

|  | Unique Analysis | Business Process-Embedded | Spreadsheet Application |
|---|---|---|---|
| Intended Frequency of Use | One-Off (One project/engagement) | Regular Routine | Heavy Massive |
| Who Can Access | Developer (Tight team) | Select personnel (Transfer over time) | Anybody (The great unwashed) |
| Number of Users | One (Tight team) | Moderate (Grows over time) | Widely distributed |
| Reliability of User Behavior | Excellent (Not shared) | High (Only bus proces owners) | Minimal (Out in the Wild) |

*Table 1: Aspects of the Use Cases*

Each column of Table 1 indicates the typical characteristics of spreadsheets for that use case. Not all spreadsheets for a use case will have all these characteristics, but they seem generally to apply. It is important to note that a given spreadsheet could appear in any of the three columns of Table 1, because use case is about the business situation, not about the spreadsheet artifact.

The alert reader will notice that risk of misuse or inadvertent damage to the source code of the spreadsheet is not included in Table 1. This is because risk is affected by the level of engineering resources (Section 4) devoted to spreadsheet development. (This is implicit in the GMS assumption that risk is appropriately managed when there is alignment between Use Case and Engineering Resources.)

The use case aspects in Table 1 will be important in Section 5 when we combine them with the engineering resource levels. We next detail the three engineering resource levels.

## 4. THE THREE ENGINEERING RESOURCE LEVELS

We now provide more detail on the engineering resources dimension of the Use Case-Engineering Resources Taxonomy in Figure 5. There are three engineering resource levels: Low, Medium, and High. We address each in turn, considering several aspects of how a spreadsheet model is resourced during development. (Disclaimer: It is our hope that these engineering resource levels and aspects encapsulate a significant fraction of important real-world spreadsheet models – and therefore are useful – but we are not quite prepared to claim that they capture every possible circumstance.)

The <u>Low</u> engineering resources level is a situation where relatively little expense and effort is devoted to the spreadsheet, and development is relatively quick. The spreadsheet is developed with little conscious regard for following any development process or meeting any standards. For example, the spreadsheet the author developed to request funding for a trip to the EuSpRIG conference; a back-of-the envelope estimate of the cost of a new product; the amount of tax-deductible interest associated with a bond that might be issued.

The <u>Medium</u> engineering resources level is a situation where relatively moderate cost, effort and time are devoted to the spreadsheet. There is some attention to process, and some level of development standards (perhaps rigorous and/or documented). To the extent that there are development standards, they are meant to ensure accuracy and reusability by trained personnel, but would be risky for extensive distribution. There is an expectation that the developer has an appropriate level of skill.

The <u>High</u> engineering resources level is a situation where relatively sizable cost, effort and time are devoted to the spreadsheet. These very high development standards build on the Medium engineering resources level, with additional work such as a sophisticated user interface that safeguards source code, vets model inputs, ensures easy future usage, etc. There is an expectation that the developer has an appropriate level of skill (extensive effort by a naïve developer is not considered High engineering resources).

Although these definitions are necessarily rough, we find them to be useful and welcome future research to refine them.

In Table 2, we describe three aspects of each engineering resource level: development standards; cost/effort; and relative time.

|        | Development Standards          | Cost/Effort | Relative Time     |
|--------|--------------------------------|-------------|-------------------|
| High   | "Unbreakable"                  | High        | Lots              |
| Medium | Formal (Rigorous, Documented)  | Moderate    | Some              |
| Low    | Informal (Non-Existent)        | Low         | Not much (Now)    |

*Table 2: Aspects of the Engineering Resource Levels*

In the next section we combine the engineering resource level aspects in Table 2 with the use case aspects in Table 1. This allows us to examine each of the nine cells in the taxonomy.

5. **THE NINE TYPES OF SPREADSHEETS IN THE USE CASE-ENGINEERING RESOURCES TAXONOMY**

This section presents the use case-resources taxonomy. There are three Use Cases, and three Resource levels, yielding a combination of nine different spreadsheet types. These are summarized in the Taxonomy shown in Figure 7. The four shaded cells in Figure 7 are situations where spreadsheet models are commonly seen in the literature and widely accepted as being desirable. The unshaded cells are situations where the literature is less informative – perhaps because these situations are less desirable – but we believe that examples of these can be found in the field.

The value of this taxonomy is three-fold. First, each cell merits its own set of development goals and guidelines. Second, the risk of error and any concomitant business loss is likely to be different for each cell; to the extent that any of the cells have not been adequately studied, this framework provides fertile ground for future research into spreadsheet development guidelines, risk assessment, and risk management. Third, the taxonomy provides a structure (see Section 6.1) for understanding how a spreadsheet can transition (with purposeful intent) or evolve (organically or by accident) over time.

|  | Unique Analysis | Bus Proc-Enabled | Sprdsht App |
|---|---|---|---|
| **High** Engineering Resources | Hobby Model | Gold-Plated | Planned App |
| **Medium** | Analytic Solution | IQASM | Incidental App |
| **Low** | Pers Prod | Field Expedient | Accidental App |

Use Case

*Figure 7: The Nine Types of Analytical Spreadsheet Models, where Shading Indicates Types Commonly Visible in the Literature*

We now examine each of the nine cells. We explain our proposed name for the type of spreadsheet in the cell. We discuss selected literature, and for the unshaded cells provide hypotheses for good and bad reasons why such spreadsheets might arise in practice. Testing these hypotheses (or developing additional hypotheses) could be a fruitful area of future research.

### 5.1 The Business Process-Enabled Use Case (Center Column)

Recall from Section 3 that the business-process enabled use case is where a spreadsheet model is used on a regular or routine basis as part of an established business process. We discuss the three types of models in this middle column of Figure 7.

**Industrial-Quality Analytical Spreadsheet Model ("IQASM")**

We start our tour of the nine types of analytical spreadsheet models in the very center cell of Figure 7. This cell combines the *Business Process-Enabled Use Case*, and has *Medium Engineering Resources*; engineering resources are in alignment with the use case. This cell is the home of the Industrial-Quality Analytical Spreadsheet Model (IQASM in Figure 7), which is a spreadsheet that satisfies six characteristics identified in GMS: suitable for efficient analysis; readable; transferable; accurate; reusable; modifiable. Notice that these characteristics are exactly aligned with the *Business Process-Enabled Use Case* attributes in Table 1. Implicit in these characteristics is *Medium Engineering Resources* (enough to satisfy the six characteristics, but less than the High Engineering Resources necessary to build an advanced user interface).

Olavson and Fry (2008) refer to spreadsheets in this cell as a "Tool", and provide a definition consistent with the six characteristics. They indicate that their group has built and handed-off many such spreadsheets. Read and Batson (1999) seem to include IQASMs as they mention (but do not emphasize) hand-over from developer to user. Based on Grossman, Mehrotra, and Özlük (2007), we believe that IQASMs are easily found in the field.

**Gold-Plated Spreadsheet Model ("Gold-Plated")**

Moving up, the top-center cell of Figure 7 combines the *Business Process-Enabled Use Case* and *High Engineering Resources*; engineering resources may be excessive for the use case. We call this situation a Gold-Plated Spreadsheet Model ("Gold-Plated" in Figure 7), which has development effort that is more than is required to fulfill its job of enabling a business process with a vetted user.

However, creating such a spreadsheet could be desirable if there was a worry that future users would not be carefully selected, or that careful hand-offs might not be sustainable; equally, such spreadsheets could represent a waste of time and money spent on overengineering unnecessary features.

**Field Expedient Spreadsheet Model ("Field Expedient")**

Moving down the center column to the bottom cell of Figure 7, we encounter the *Business Process-Enabled Use Case* and *Low Engineering Resources*; engineering resources are insufficient for the use case. Here resides what we call the Field Expedient Spreadsheet Model ("Field Expedient" in Figure 7), which is a spreadsheet that is in regular use in a business process but has not been developed sufficiently to make it an IQASM that is most suitable for that role. (We are not sure that "field expedient" is the best name, but it has the benefit of positivity.)

Having such a spreadsheet could be a smart business move, for example a business situation where a fast handoff of a useful but lesser-developed model (accepting that its use might be problematic since it is designed for use only by the developer) is deemed preferable to not having a model at all – that is, it's a risk, but a smart risk. Equally, it could arise due to thoughtless, naïve, or careless practices. We hypothesize that some number of spreadsheet errors and problems can be traced to Field Expedient spreadsheet models.

### 5.2 The Unique Analysis Use Case (Left Column)

Recall from Section 3 that the unique analysis use case is where a spreadsheet model is built to undertake a single calculation or to provide a model used for analysis in a situation that is limited in scope or time. We discuss the three types of models in this left-hand column of Figure 7.

**Personal Productivity Spreadsheet Model ("Pers Prod")**

A Personal Productivity Spreadsheet Model ("Pers Prod" in the bottom left cell of Figure 7) is the common situation of a minor spreadsheet, which might be the proverbial "quick and dirty" spreadsheet model, which is intended to be used once for something of interest (*Unique Analysis Use Case*), that gets little developmental time and attention (*Low Engineering Resources*).

We hypothesize that the overwhelming majority of the world's analytical spreadsheet models fall into this category.

**Analytical Solution Spreadsheet Model ("Analytical Solution")**

Moving up the *Unique Analysis Use Case* column of Figure 7, we encounter the center row, *Medium Engineering Resources*. We refer to this cell as the Analytical Solution Spreadsheet Model ("Analytical Solution" in Figure 7), a term originated by Olavson and Fry (2008). This important type of spreadsheet model cannot be resolved in the three-type GMS taxonomy.

We suggest that a reasonable criteria for inclusion in this cell is a subset of the six characteristics for an Industrial Quality Analytical Spreadsheet Model (see section 5.1 above). The use case is one-off analysis by the spreadsheet developer (or tight team), hence we remove the characteristics associated with handing off the spreadsheet to a less sophisticated business process owner, and can skip the investment for the spreadsheet to be transferable, modifiable, and readable (in the sense of easily readable by a person not familiar with the spreadsheet's development; see GMS). This yields three characteristics that might define the Analytical Solution: suitable for efficient analysis; accurate; reusable. (We revisit this in Section 6 below.)

The Analytical Solution type is vitally important in certain business contexts. Olavson and Fry view them as central to their practice at Hewlett-Packard. The work in section 2.1 of Tennent and Friend (2005) is an Analytical Solution. Read and Batson (1999) seem to focus on Analytical Solutions, but there work is somewhat casual with regard to use case. The work of Swan (2008), and some or all of the models made by spreadsheet manufactories (discussed further in Section 6) seem to be Analytical Solutions (see Grossman and Özlük 2010).

**Hobby Spreadsheet Model ("Hobby Model")**

The top row of the *Unique Analysis Use Case* column of Figure 7 embraces *High Engineering Resources*. We refer to this cell is as the Hobby Spreadsheet Model ("Hobby Model" in Figure 7). These are spreadsheets that are suitable for analytical use by their developer, that have received further investment to make them difficult to damage or use improperly. Since they are being used by their creator, such investment seems excessive, leading to the "hobby" moniker. However, there may be valid reasons for this investment, such as making provision for possible future unplanned evolution to the business process-enabled use case in Figure 7. (Such evolutions are discussed in Section 6.2.)

**5.3 The Spreadsheet Application Use Case (Right Column)**

Recall from Section 3 that the spreadsheet application use case is where a spreadsheet model is written by a developer and distributed to other people (with an unspecified level of intentionality and planning), resulting in usage of the spreadsheet by multiple users, with the possibility of user modifications of the spreadsheet if that is not prevented by application of engineering resources. We discuss the three types of models in this right-hand column of Figure 7.

**Planned Application Spreadsheet Model ("Planned App")**

The top row of the *Spreadsheet Application Use Case* column of Figure 7 meets *High Engineering Resources*. We refer to this cell is as the Planned Application Spreadsheet Model ("Planned App" in Figure 7). This cell contains spreadsheet models that are purposefully built for distribution to distributed users.

One would expect spreadsheets of this type to have sufficient engineering investment so that it is reasonable to send a copy to relatively unreliable personnel, including provisions to prevent users from editing the source code, and trapping input values that will cause errors or spurious results. Several examples of planned application spreadsheet models can be found in Grossman, Mehrotra, and Özlük (2007), hence it is likely that these are easily found in the field.

**Incidental Application Spreadsheet Model ("Incidental App")**

The Incidental Application Spreadsheet Model ("Incidental App" in Figure 7) arises when a spreadsheet that serves the *Spreadsheet Application Use Case* and has received only *Medium Engineering Resources*. Although the engineering resources are comparable to those for the IQASM cell (to the left in Figure 7), these spreadsheets tend to lack provision for source code protection and input error trapping, and therefore carry a certain level of risk when used by unreliable users.

We are not sure what the literature has to say about these. We have heard anecdotes of such spreadsheets rapidly forking source code (making them difficult or impossible to maintain) and generating spurious results that are blamed on the original developer. Such spreadsheets might arise through naivete, for example, sending an IQASM to many people. But it is not unlikely that there are times in a business setting where the benefits of distribution outweigh the expense and risk.

**Accidental Application Spreadsheet Model ("Accidental App")**

The Accidental Application Spreadsheet Model ("Accidental App" in Figure 7) arises when a spreadsheet serving the *Spreadsheet Application Use Case* has received only *Low Engineering Resources*. These spreadsheets, which have little or nothing in the way of thoughtful construction or conformance to any standards, and are suitable only for use by the developer, seem (we are being polite) poised to generate exciting results as multiple unreliable users have adventures with them.

The caveats that apply to Incidental Application Spreadsheet Models above apply even more strongly here. We are optimistic that a determined researcher could find a reason where such a spreadsheet type is desirable, but we have been unable to articulate any. This class of spreadsheets seems to carry with it high risk of error, including inadvertent or incorrect changes to source code, as well as inappropriate usage. We are not aware of any literature on this class of spreadsheet, but they undoubtedly exist in the wild.

## 6. CONCLUSIONS AND AREAS FOR FUTURE RESEARCH

We present a new taxonomy for understanding analytical spreadsheet models, which are a domain of spreadsheets of particular practical importance. Nine types of spreadsheets are identified, whereas the previous GMS taxonomy had only three.

The use case-engineering resources taxonomy is grounded in three use cases: spreadsheet models meant for unique analysis; business-processed enabled spreadsheet models that are embedded in a regular or routine activity; and spreadsheet applications that are handed off to multitudinous users. The taxonomy incorporates three level of spreadsheet engineering resources applied to models for each use case: low resources where little attention is paid to practices and standards; medium resources where the spreadsheet model is made suitable for regular use; and high resources where the spreadsheet model has safeguards to the source code and inputs rendering it difficult to "break".

We briefly discuss or sketch each of the nine types of analytical spreadsheet model in the taxonomy. We indicate the type of spreadsheets one might see for that type, and present formal inclusion criteria (with varying degrees of confidence) for some of them. We describe how these spreadsheets can be found in the literature. We share our perception of how easily they might be found in the field. We speculate on reasons (good and bad) that such spreadsheet models might be created in the rough-and-tumble of business, and discuss their relative riskiness and utility.

This taxonomy is best thought of as a coherent set of ideas that will be helpful to strengthening our understanding of existing spreadsheet development recommendations; the way that spreadsheets change over time; and an individual's or organization's decision-making around investing (or not) time and energy into spreadsheet models. There is much research that could be performed to test, explore, and deepen our understanding of this new use case-engineering resources taxonomy.

**6.1 Testing and Refining the Use Case-Engineering Resources Taxonomy**

We recognize that the domain of analytical spreadsheet models is very large, and it is highly ambitious to define a framework that can define the entire space. We are optimistic that this taxonomy is useful. There is opportunity for further work on all aspects of the taxonomy. The engineering resource levels would benefit from better definition. The use case could be more deeply explored. All cells should have a set of criteria for membership, and their names might merit refinement.

The five cells that are not shaded in Figure 7 are less visible in the literature, and for reasons discussed in Section 5 are inherently problematic. These cells have potential for fruitful research.

For the IQASM cell (the center of Figure 7) we use the criteria from GMS. We suggest a set of criteria for Analytical Solution based on a reduction of the criteria for IQASM. Both of these sets of criteria should be tested against spreadsheets in the literature and in the field.

It would be desirable systematically to examine the field research on spreadsheets in light of this taxonomy, including Croll (2007) and Grossman, Mehrotra, and Özlük (2007). This taxonomy could usefully be examined through the lens provided by the unusually detailed FAST Standard (FAST 2019). It would also be helpful to examine this taxonomy in light of the largely-unpublished, rich, practical expertise of spreadsheet manufactories such as Operis Group, Ltd., F1F9, Modano (which seems to have absorbed and ceased support for SSRB/BPM), and undoubtedly many other organizations, and also in light of the guidelines and wisdom of practice available from e.g., ICAEW, F1F9's delightful Resources page (F1F9 2023), etc.

It would be interesting to consider errors in analytical spreadsheet models in light of this taxonomy. Are there types of spreadsheets in this taxonomy that are particularly prone, or less prone, to error?

**6.2 The Evolution and Transition of Spreadsheets Over Time**

The development guidelines in the literature tend to be static. That is, they provide recommendations for a spreadsheet in a particular domain as a one-off development project. The framework has potential for to help us understand how spreadsheets change over time, and crafting recommendations for how they should evolve.

We suggest that there might be two ways that a spreadsheet might over time change its type in Figure 7. The first is purposefully, with due care, planning, and investment. We refer to this as *transition* from one type to another. The second is organically (or accidentally), where over time a spreadsheet shifts from one type to another type, without much in the way of forethought or purposeful investment. We refer to this change as *evolution* from one type to another.

There seems to be little literature on managing the rightward transition of a spreadsheet from Unique Analysis to Business Process-Embedded (which is a common transition), to Application (which we hypothesize is common). We hypothesize that in terms of Figure 7, a transition will tend to be a move up a row in the same column, or diagonally up and to the right, whereas an evolution would tend be a move to the right of the same row, reflecting a spreadsheet being distributed more broadly without additional engineering resources.

Guidelines for making smart engineering investments, with an eye to purposeful transition rather than organic/accidental evolution seem desirable. For example, sections 5.1 and 5.2 suggest that the transition would require engineering investment in modifiability, transferability, and readability. Indeed, Olavson and Fry (2008) discuss in detail the transition (to the right) from Analytical Solution to Industrial-Quality Analytical Spreadsheet Model. It would be valuable to generalize their insights beyond their domain of supply chain analytical models at Hewlett-Packard.

As another example, if an analyst has a Personal Productivity Spreadsheet (Unique Analysis Use Case & Low Engineering Resources) and realizes that it is becoming embedded in a routine business process (Use Case changes to Business Process-Enabled), what guidance could be provided for extending the spreadsheet up and to the right to become an Industrial-Quality Spreadsheet Model, hence avoiding the more risky outcome of becoming a Field Expedient?

## 7. ACKNOWLEDGMENT

We thank Professor Kourosh Dadgar for his helpful comments on a draft of this paper. We are grateful to two anonymous referees whose comments improved the paper. All errors of omission or commission are the responsibility of the authors.